\documentstyle[12pt]{article}
\newlength{\dinwidth}
\newlength{\dinmargin}
\newcommand{\be}{\begin{equation}}
\newcommand{\ee}{\end{equation}}
\newcommand{\nin}{\noindent}
\newcommand{\CC}{{\mbox{C\hspace{-0.60em}I\hspace{0.00em} }}}
\newcommand{\II}{{\mbox{I\hspace{-0.51em}I\hspace{0.00em} }}}
\newcommand{\NN}{{\mbox{N\hspace{-0.90em}I\hspace{0.30em} }}}
\newcommand{\RR}{{\mbox{R\hspace{-0.90em}I\hspace{0.10em} }}}

\newcommand{\la}{{\lambda}}                     
\newcommand{\dsx}{d \mbox{\boldmath $x$}}       
\newcommand{\sx}{\mbox{\footnotesize{\boldmath $x$}}}
\newcommand{\Oo}{{\cal O}}                      
\newcommand{\Op}{{\cal O}'}                     
\newcommand{\Oi}[1]{{\cal O}_{#1}}              
\newcommand{\AO}{{\cal A}({\cal O})}            
\newcommand{\AOi}[1]{{\cal A}({\cal O}_{#1})}
\newcommand{\Al}{{\cal A}}                      
\newcommand{\Li}{{\cal L}}                      
\newcommand{\Co}{{\cal C}}                      
\newcommand{\AlO}{{\cal A}_{\lambda} ({\cal O})}
\newcommand{\Alu}{\underline{{\cal A}}}         
\newcommand{\AOu}{\underline{{\cal A}} ({\cal O})}
\newcommand{\AOiu}[1]{\underline{{\cal A}} ({\cal O}_{#1})}
\newcommand{\Au}{\underline{A}}                 
\newcommand{\Aul}{{\underline{A}}_{\lambda}}    
\newcommand{\Abs}[1]{ \left| #1 \right|}
\newcommand{\ABS}[1]{ \left\| #1 \right\|}
\newcommand{\Pg}{{\cal P}_{+}^{\uparrow}}       
\newcommand{\Lx}{( \Lambda , x )}               
\newcommand{\ax}{{\alpha}_{x}}                  
\newcommand{\aL}{{\alpha}_{\Lambda}}            
\newcommand{\aLx}{{\alpha}_{\Lambda , x}}       
\newcommand{\Ux}{U(x)}                          
\newcommand{\Uox}{U_{\omega}(x)}                
\newcommand{\UoL}{U_{\omega}(\Lambda)}          
\newcommand{\axu}{{\underline{\alpha}}_x}       
\newcommand{\aLu}{{\underline{\alpha}}_{\Lambda}}
\newcommand{\dmu}{{\underline{\delta}}_{\, \mu}}
\newcommand{\oz}{{\omega}_{0}}                  
\newcommand{\Ho}{{\cal H}_{\omega}}             
\newcommand{\Ooo}{{{\Omega}_{\omega}}}          
\newcommand{\po}{{\pi}_{\omega}}                
\newcommand{\oziu}{{\underline{\omega}}_{{0}, \iota}}
\newcommand{\ou}{{\underline{\omega}}}          
\newcommand{\pu}{{\underline{\pi}}}             
\newcommand{\bra}[1]{\langle {\mbox{$#1$}} |}   
\newcommand{\ket}[1]{| {\mbox{$#1$}} \rangle}   
\newcommand{\ketl}[1]{{|{\mbox{$#1$}} \rangle}_{\lambda}}
\newcommand{\bracket}[2]{\langle {\mbox{$#1$}} | {\mbox{$#2$}} \rangle}
\title{{\bf On the Manifestations of Particles}\thanks{Talk given at
the "Workshop on Mathematical Physics Towards
the 21st Century", Beer Sheva, 14-19 March 1993}
}
\author{Detlev Buchholz \\
II. Institut f\"ur Theoretische Physik,
Universit\"at Hamburg \\
D-22761 Hamburg, Federal Republic of Germany}
\date{}
\begin{document}
\maketitle
\begin{abstract}
The mathematical description of stable
particle-like systems appearing in relativistic quantum field theory
at large, respectively small scales or non-zero temperatures is
discussed.
\end{abstract}
\section{Introduction}

Stable particle-like systems are met in high energy physics in very
different settings: elementary particles, as well as their bound
states, are observed at asymptotic times in collision processes. At the
other end of the scale, i.e. at short distances, particle-like
structures confined in hadronic matter (partons) appear in
deep inelastic processes. Proceeding to thermal states, they may
also exist as individual constituents of a hot plasma (quarks,
gluons). It is a somewhat embarassing fact that for most of these
fundamental elementary systems a fully satisfactory mathematical
description is not at hand.

In the conventional treatment, originating from E. Wigner's
pioneering analysis \cite{W}, particles are identified with vector
states in irreducible representations of the Poincar\'e group, resp.
its covering group. This characterization of particles has proved
to be extremely useful: it led to an understanding of the role
of mass and spin, it is a key ingredient in the construction of
field theoretic models, entering in the renormalization procedure,
and it is the basis for the physical interpretation of the theory in
terms of collision states. These applications reveal the value
of a sound {\em a priori} concept of particle within the theoretical
setting.

Wigner's particle concept applies only to a restricted class of
elementary systems, however, a well known counter-example being
particles carrying electric charge. The states of such particles
always contain infinitely many low energy photons and therefore
cannot be described by vectors in some Poincar\'e invariant
superselection sector of the physical Hilbert space; in
fact, neither mass nor spin can sharply be defined for these states
\cite{FMS1, B1}. The inappropriate Wigner concept of particle is
at the origin of the infrared problems in the early treatments
of quantum electrodynamics and led B. Schroer to the notion of
infraparticle \cite{Sch}. Although one has learned in the
meantime how to handle these difficulties, a clear-cut mathematical
characterization of electrically charged particles has not yet
emerged. Summarizing the folklore, these particles are regarded
as certain limit cases of Wigner type particles, which are
accompanied by photon clouds (cf. \cite{FMS2, B2} for a rigorous
formulation of this idea). This picture suffices in many
applications for a consistent interpretation of theoretical
calculations, e.g. of collision cross sections. But it somewhat
obscures the genuine nature of the underlying physical systems.

In the case of partons the situation is even less satisfactory.
These entities made their first appearance in an ingenious
interpretation of experimental data \cite{Fe}, relying on Wigner's
particle concept. Their theoretical description has later been
formalized in the perturbative treatment of quantum chromodynamics,
which also relies on that concept through the use of Feynman rules.
In this way the existence of partons has now been firmly
established, but it is also clear that the particle concept
used has only a tentative meaning since partons are (expected
to be) confined and physical states of these entities do not
exist. Again one pays for the inappropriate particle concept by
tantalizing infrared problems in the perturbative setting and the need
to introduce in an {\em ad hoc} way phenomenological fragmentation
functions which are tailored to convert the perturbative results into
physically significant statements. These difficulties may in
principle be avoided by considering only physical states, as e.g. in
lattice gauge theory. But then one loses contact with the simple
parton picture, and the problem of a description of the parton
content of states reappears.

Problems of still another kind arise if one wants to describe
the particle features of thermal states. Thinking of the
deconfinement transition in quantum chromodynamics, for example,
it would be of interest to characterize the constituents of such
states. Of similar physical interest are the quasiparticles
\cite{L}, i.e. collective excitations of the constituents which
frequently exhibit particle properties as well.
The constituents of a thermal state may be expected to have
properties which do not deviate too much from those of an
elementary particle, at least in the case of low densities.
Nevertheless, their mathematical description causes problems
because of the influence of the thermal background. In fact,
neither the constituents, nor the quasiparticles can obey a
sharp dispersion law if there is interaction \cite{NRT}. This
fact makes it difficult to distinguish them from other modes,
and again one runs into inconsistencies if one imposes from
the outset a wrong particle concept \cite{La}.

It is the aim of this article to report on attempts to overcome
these conceptual difficulties. The results obtained so far are
not yet complete, but they shed new light on the problems and seem
to be a solid basis for further studies. Thus, in accord with
the idea of this workshop, this contribution intends to be both,
a status report and a bit of science fiction.

The article is organized as follows: Sect. 2 contains an outline
of algebraic quantum field theory, which is the natural framework
for particle analysis. The particle structures which appear
in this setting at asymptotic times are described in Sect. 3.
These results are fairly complete and lead to a unified treatment
of elementary particles and infraparticles in terms of weights
\cite{BPS}. Moreover, they provide a general method by which the
asymptotic particle content can be extracted from any state.

The particle structures appearing in the short distance limit are
the theme of Sect. 4. For their characterization the concept of a
scaling algebra is introduced which provides an algebraic version of
the renormalization group and allows a systematic discussion
of the scaling limit of any quantum field theory \cite{BV}.
It turns out that this limit can always be interpreted in terms of
vacuum states on the scaling algebra. The particle content of the
theory in the short distance limit can thus be determined by a
particle analysis of the local excitations of these vacua, as
indicated above, and one may therefore hope for a complete
classification of the pertinent structures.

Section 5 is dedicated to the particle aspects of thermal
equilibrium states. For the investigation of the particle structures
in the scaling limit one can apply the method of the scaling
algebra, yielding the same results as before. But if one proceeds
to larger scales, the analysis becomes more difficult since the
specific dynamical properties of the thermal background come
into play. Nevertheless it appears to be possible to distinguish
the constituents of thermal states in precise mathematical terms
\cite{BB1}. A clear-cut mathematical characterization of quasiparticles
remains, however, as an intriguing problem.

The article closes with some "concluding ruminations".
\section{The framework}

Before entering into details, it may be helpful to point to the
similarities of our approach and the strategy of Wigner. Wigner took
as the starting point of his analysis the principles of special
relativity, which are encoded in the properties of the space-time
symmetry group, the Poincar\'e group. Since symmetries act on physical
states by automorphisms, it appeared to be natural to assume that
particles, being the most elementary systems, give rise to irreducible
representations of the Poincar\'e group, respectively its covering
group.\footnote{At the time of its invention this assumption
must have seemed to be exceedingly general. But today we see its
weak point: symmetries can be spontaneously broken so that their
action on states gives rise to incoherent systems, belonging to
different superselection sectors. It then makes no sense to
classify the states according to the irreducible representations
of the symmetry group} This idea of a particle then led naturally
to the problem of classifying these representations \cite{W}.

In the present approach we start from the principle of locality,
which is encoded in algebraic relations between observables. The
notion of locality is of vital importance in our characterization
of particles, where we appeal to the heuristic picture of a particle
as a stable, singly localized entity \cite{E}. This picture will lead
us to a comprehensive description of particles in terms of specific
states (resp. weights) on the algebra of observables, depending
on the respective physical sitution. Thus, like Wigner, we characterize
particles by a condition of simplicity, and we are then led to the
classification of the corresponding representations of the
observables, in complete analogy to his approach.

As already indicated, we will make use of the framework of
algebraic quantum field theory \cite{H} which allows us to discuss
the particle aspects of relativistic quantum field theories in full
generality. For the convenience of those readers who are not
familiar with this setting, we briefly list the relevant
assumptions and add a few comments.

1.  {\em (Locality)} We suppose
that the local observables of the underlying theory generate a net
of local algebras over $d$-dimensional Minkowski space $\RR^d$, i.e. an
inclusion preserving map
\be \Oo \rightarrow \AO \ee
from the set of open, bounded regions
$\Oo$
in Minkowski space to unital C$^*$-algebras
$\AO$.
We will think of
$\AO$
as the algebra generated by observables which can be measured in
$\Oo$.
The algebra generated by all local algebras
$\AO$
(as a C$^*$-inductive limit) will be denoted by
$\Al$.
The net is supposed to satisfy the principle of locality (Einstein
causality), i.e.
\be \AOi{1} \subset \AOi{2}' \: \mbox{if} \: \Oi{1} \subset \Oi{2}',\ee
where
$\Op$
denotes the spacelike complement of
$\Oo$
and
$\AO '$
the set of operators in
$\Al$
which commute with all operators in
$\AO$.
We note that this abstract algebraic setting is quite natural for our
purposes since we are interested
in physical systems whose Hilbert space description has yet to be
determined.

2. {\em (Covariance)} The Poincar\'e group
$\Pg$
is represented by automorphisms of the net. Thus for each
$\Lx \in \Pg$
there is an
$\aLx \in \mbox{Aut} \Al$
such that, in an obvious notation,
\be \aLx ( \AO ) = \Al ( \Lambda \Oo + x ) \ee
for any region
$\Oo$.
We amend this fundamental postulate by a continuity condition,
i.e. we assume that for any
$A \in \Al$
the function
$\Lx \rightarrow \aLx (A)$
is strongly continuous. This is no loss of generality since
these observables suffice to separate the states of physical interest.

3. {\em (States)} Physical states are described by positive, linear
and normalized functionals
$\omega$
on
$\Al$.
By the GNS-construction, any state
$\omega$
gives rise to a representation
$\po$
of
$\Al$
on a Hilbert space
$\Ho$,
and there exists a cyclic vector\footnote{This means that
$\po ( \Al ) {\Omega}_{\omega}$
is dense in
$\Ho$}
$\Ooo \in \Ho$
such that
\be \omega (A) \ = \ ( \Ooo, \po (A) \Ooo ), \ A \in \Al. \ee
Since the dual space of
$\Al$
contains an abundance of states corresponding to very different
physical situations (few and many body systems, equilibrium and
non-equilibrium systems, etc.), one has to select the states of
actual interest by additional conditions.

It is a characteristic feature of few body systems that they have finite
total energy and momentum. This suggests to characterize the
corresponding states
$\omega$
by the condition \cite{Bo} that the space-time translations
$x$
act on the GNS-space
$\Ho$
by a continuous unitary representation
$\Uox$ which
satisfies the relativistic spectrum condition (i.e. the
generators have joint spectrum in the closed forward lightcone
$\overline{V}_+$)
and implements the translations of observables,
\be \Uox \po (A) {\Uox}^{-1} \ = \ \po ( \ax (A)),  \, A  \in \Al.
\ee
As a matter of fact, under these circumstances there exists a
unique choice of the representation
$\Uox$
with Lorentz invariant spectrum, as a consequence of locality
\cite{BoB}. States
$\omega$
and representations
$\po$
with these properties are said to have positive energy. If a
positive energy state
$\omega$
is invariant under translations, i.e.
$\omega ( \ax (A) ) = \omega (A)$
for all
$x$
and
$A \in \Al$,
its GNS-vector
$\Ooo$
is invariant under the action of
$\Uox$
and may thus be interpreted as vacuum.

In case that also the Lorentz transformations
$\Lambda$
act on
$\Ho$
by a continuous unitary representation
$\Lambda \rightarrow \UoL$,
and
$\UoL \po (A) {\UoL}^{-1} = \po ( \aL (A) ), A \in \Al$,
we speak of Poincar\'e covariant representations. It is reasonable to
assume that states
$\omega$,
representing the vacuum, are invariant under Lorentz
transformations, thereby giving rise to Poincar\'e covariant
representations. But, as was pointed out in the Introduction, the
condition of full Poincar\'e covariance for non-vacuum states is
in general too restrictive, as it excludes from the outset states
carrying electric charge.

We will also consider states
$\omega$
on
$\Al$
which describe systems in thermal equilibrium. These states are
characterized by the KMS-condition, which is the proper
description of Gibbs ensembles in the thermodynamic limit
\cite{HHW,H}. In the formulation of this condition there enters a
distinguished time axis, which is fixed by the rest frame of
the heat bath. Denoting by $t$ the translations along this axis, the
KMS-condition says that
$\omega$
is a thermal equilibrium state at inverse temperature
$\beta > 0$
if the following holds: for any given
$A,B \in \Al$
the functions
$t \rightarrow \omega ( A \alpha_{t} (B) )$
and
$t \rightarrow \omega ( \alpha_{t} (B) A )$
are continuous boundary values of some analytic function in the
strip
$\{z \in \CC : 0 < \mbox{Im} z < \beta \}$,
at its lower and upper rim, respectively.
\section{Particles and infraparticles}

States
$\omega$
of positive energy are expected to describe at asymptotic times
configurations of a few stable particles, possibly accompanied
by clouds of low energy massless particles. Following this idea
and the conception of a particle as some well-localized entity,
it seems natural to analyze the asymptotic particle aspects of
$\omega$
by studying the properties of the expectation values
$\omega ( \ax (A) ), A \in \Al$,
for large timelike $x$. The dominant contributions should originate
from a vacuum state
$\oz$,
\be \lim_{x} \omega ( \ax (A) ) = \oz (A), \ A \in \Al, \ee
since the probability for finding some deviation from the vacuum
in a finite space-time volume about $x$ (fixed by the localizaton
properties of the respective observable $A$) should tend to $0$
in a state of finite energy because of dispersive effects
("spreading of wavepackets").\footnote{This argument
is actually not quite correct since the vacuum may be "slowly
varying" at infinity, cf. \cite{BW}. But one can show within the
present general setting that the family of states
$\omega \circ \ax$
always has weak *-limit points which are vacuum states} We are here
primarily interested
in the local deviations from the vacuum, however. Following Araki
and Haag \cite{AH}, we therefore proceed to a subalgebra
$\Co \subset \Al$
of observables which are insensitive to the vacuum.

The algebra $\Co$ is constructed as follows: we first consider the
vector space
${\Li}_0 \subset \Al$
of almost local operators\footnote{An operator
$B \in \Al$
is said to be almost local if there exists an approximating
sequence
$B_r \in \AOi{r}, \ \Oi{r}$
being a family of concentric balls of radius $r$, such that
$\ABS{B - B_r}$
tends to
$0$
for large $r$ faster than any inverse power of $r$}
$L_0$
such that the functions
$\Lx \rightarrow \aLx (L_0)$
are C$^{\infty}$
and have, with respect to $x$, Fourier transforms with support in
some arbitrary compact subset
$\Gamma$
in the complement of the closed forward lightcone
$\overline{V^+}$
(they "transfer" energy-momentum
$\Gamma$).
Such operators are easily obtained by regularization of (almost)
local operators
$A \in \Al$
with respect to the action of the Poincar\'e group,
\be L_0 = \int    d \mu \Lx    f \Lx \aLx (A). \ee
Here
$\mu$
is the Haar-measure on $\Pg$ and $f$ is any test function whose Fourier
transform with respect to $x$ has support in some
$\Gamma$.
By left multiplication of the elements of
${\Li}_0$
with arbitrary elements from
$\Al$
and linear combinations we obtain a (non-closed) left ideal
$\Li \subset \Al$
which is invariant under the action of the Poincar\'e group. We call
$\Li$
the left ideal of localizing operators for reasons which will
become clear later. Because of the specific energy-momentum transfer
$\Gamma$
of the operators in
${\Li}_0$,
any operator
$L \in \Li$
annihilates the vacuum state(s) and, as a matter of fact, all positive
energy states of sufficiently small energy (depending on $L$).

We then define the algebra
$\Co$
as the linear span of all operators of the form
$L^* L$, $L \in \Li $,
i.e.
$\Co = {\Li}^* \Li$
in shorthand. The algebra
$\Co$
is smaller than the "algebra of detectors" introduced in \cite{AH}.
But it separates all positive energy states which are
orthogonal to the vacuum and is therefore big enough for our
purposes.

We can now proceed with our asymptotic analysis of
positive energy states. Since the operators
$C \in \Co$
annihilate the vacuum, the at asymptotic times leading
contributions to the expectation values
$\omega ( \ax (C) ), C \in \Co$,
should be due to single stable particles, triggering $C$.
Contributions from multiple excitations of the vacuum may be
expected to appear only in next to leading order since the
probability for finding more than one particle in a space-time
region of fixed size should decay substantially faster with time.
As far as the single particle contributions in the functions
$x \rightarrow \omega ( \ax (C) )$ are concerned, they
will tend to zero at large times because the particles can be
anywhere, one loses control of their localization. This picture
suggests to integrate these functions over spacelike surfaces and
thereby to compensate the effect of dispersion.

At this point there arises a problem: do the integrals
$\int \dsx \,  \omega ({\alpha}_{t, \sx} (C) ), C \in \Co$
(where we have introduced proper coordinates
$x = (t, {\mbox{\boldmath $x$}})$)
exist? An affirmative answer, which may be regarded as a first
confirmation of our heuristic picture, follows from the subsequent
lemma \cite{B3}. Its proof is based on a harmonic analysis of the
space-time translations
${\alpha}_{\, {\RR}^d} $.\footnote{The general harmonic
analysis of automorphism
groups \cite{A} seems still in its initial stages. It would
be useful to have a classification of their possible
spectral properties, in analogy to the notion of measure classes
in the case of unitary groups}

\vspace{2mm}
\noindent {\bf Lemma 3.1:}  {\em Let
$E( \cdot )$
be the spectral resolution of the generators of space-time
translations
$\Ux$
in any positive energy representation
$\pi$
of
$\Al$.
Then, for any
$C \in \Co$
and any compact subset
$\Delta$
of the spectrum,
$$ \limsup_{r \rightarrow \infty} ||  {E ( \Delta )
\int_{\Abs{\bf x} < r} \dsx \,  \pi ( {\alpha}_{\sx} (C) )
   E( \Delta )} || < \infty.$$ }

Thus if
$\omega$
is a positive energy state whose spectral support (with respect
to the generators of the space-time translations in its
GNS-representation) is contained in some compact set
$\Delta$,
we see from the lemma that the integrals in question are
well-defined and uniformly bounded in $t$ (since the spectral
support of
$\omega$
is invariant under time translations). In fact, using the lemma,
one can equip
$\Co$
with a directed set of seminorms, and the positive linear
functionals
${\rho}_t , t \in \RR$
on
$\Co$
given by
\be {\rho}_t (C) = \int \dsx \,  \omega ( {\alpha}_{t, \sx} (C) ), \
C \in \Co \ee
form an equicontinuous family with respect to the resulting topology.
Thus there exist weak *-limit points if $t$ approaches
$\pm \infty$,
respectively. Actually, one would expect that in physically
reasonable theories the functionals
${\rho}_t$
converge since the asymptotic particle configurations should
stabilize. This question is closely related to the problem of
asymptotic completeness, cf. below. In order to deal with possible
oscillations of asymptotic particle configurations, we proceed
to time averages of the functionals
${\rho}_t$,
\be \overline{{\rho}_t} = \frac{1}{\la (t)}
\int_{t}^{t + \la (t)}    dt' {\rho}_{t'}, \ee
where
$\la$
is some positive, slowly increasing function. According to the
preceding discussion, we expect that the limit points of the
functionals
$\overline{{\rho}_t}$
contain the relevant information about the asymptotic particle
content of
$\omega$.

\vspace{2mm}
\nin {\em Working hypothesis:} The limit points
$\{ {\sigma}_{\pm} \}$
of the functionals
$\overline{{\rho}_t}, t \rightarrow \pm \infty $
describe the outgoing, respectively incoming stable particle
content of the state
$\omega$.

\vspace{2mm}
The problem of describing particles appearing at asymptotic times
has thus turned into a problem of mathematical analysis and
interpretation of the structure of the functionals
${\sigma}_{\pm}$.
Such an analysis has been carried out in cooperation with
M. Porrmann \cite{BP}, cf. also \cite{BPS}, and we give here
an account of some of the results. Since in this discission
the time direction plays no role, we omit in the following
the index
$\pm$
in
${\sigma}_{\pm}$.

It is apparent from their definition that the functionals
$\sigma$
on
$\Co$
are positive and linear. (But they are not normalizable; note that
$\Co$
does not contain a unit.) Thus any
$\sigma$
induces a positive semi-definite scalar product on the left ideal
$\Li$.
Proceeding as in the GNS-construction, one obtains, by taking
quotients with respect to elements of zero norm and completion,
a Hilbert space
${\cal H}$
and a linear map
$\ket{\cdot} : \Li \rightarrow {\cal H}$
with dense range, such that
\be \bracket{L_1}{L_2} = \sigma ( L_{1}^{*} L_2 ) \: \ L_1, L_2 \in \Li.
\ee
Here
$\bra{\cdot}$
denotes of course the corresponding antilinear map of
$\Li$
into the dual space of
${\cal H}$.
As suggested by the notation, it may be helpful to think of
$\bra{\cdot}$ and $\ket{\cdot}$
as improper Dirac bra and ket vectors, respectively, which
become normalizable after application of the elements of
$\Li$.
Only in special cases, they may also be identified with improper
states in a rigged Hilbert space formalism. Some important
properties of the maps
$\ket{\cdot}$
are summarized in the

\vspace{2mm}
\nin {\bf Proposition 3.2:} {\em
i) The operators
$\pi (A), A \in \Al$
fixed by
$$ \pi (A) \ket{L} \doteq \ket{AL}, \ L \in \Li$$
are well-defined and bounded. The corresponding map
$\pi : \Al \rightarrow {\cal B}({\cal H})$
defines a *-representation of
$\Al$. \hfill \\
ii) The functions
$\Lx \rightarrow \ket{\aLx (L)}, \ L \in {\Li}_0 $
are C$^{\, \infty}$.
(${\Li}_0$
is the generating space of
$\Li$.)  \hfill \\
iii) The operators $U(x)$ fixed by
$U(x) \ket{L} \doteq \ket{\ax (L)}, \ L \in \Li$
define a continuous unitary representation
$x \rightarrow U(x)$
of space-time translations on
${\cal H}$.
The spectrum of the generators of $U(x)$ is contained in some (shifted)
light cone
$\overline{V}_+ + q$. }

We note that the vector $q$ in the latter statement depends on the
spectral support of the underlying state
$\omega$.

\vspace{2mm}
\nin {\em Definition:} A linear map
$\ket{\cdot}$
of
$\Li$
into some Hilbert space
${\cal H}$
with dense range is called a {\em particle weight} if it has the
properties given in the Proposition.

\vspace{2mm}
The representations
$\pi$
of
$\Al$
induced by the particle weights constructed above
will be in general highly reducible. (Note that in the construction
of $\ket{\cdot}$
there appeared direct integrals of states.) Since elementary systems,
such as particles, should give rise to irreducible representations of
$\Al$,
it is natural to pose the question of whether
$\ket{\cdot}$
can be decomposed into pure components. An affirmative answer is given
in the following statement.

\vspace{2mm}
\nin {\bf Proposition 3.3:} {\em  Let
$\ket{\cdot}$
be a particle weight. There exists a measure space
$( \Sigma , \mu ) $
and a measurable family of pure particle weights
$\ketl{\cdot}, \la \in \Sigma$
(inducing irreducible representations of
$\Al$)
such that there holds the direct integral decomposition
\be \ket{\cdot} = \int    d \mu ( \la )    \ketl{\cdot}. \ee
}

\nin {\em Remark:} In this form the statement holds actually only
if
$\Al$
is separable. We refrain from discussing here the complications
arising in the general case.

This result tells us that it is possible to resolve the asymptotic
particle structure described by
$\sigma$
into components with definite superselection quantum numbers
(charges). More can be said, but before we proceed we have to discuss
some technicality.

\vspace{2mm}
\nin {\em Definition:} A particle weight
$\ket{\cdot}$
is said to be regular if
$\sum_{i=1}^{n} \ket{{L}_{i}^* {L}_{i}} = 0$
implies
$\ket{{L}_i} = 0, {L}_i \in \Li, i=1 \dots n$.
(Note that this implication would trivially hold if
$\Li$
would contain a unit.)

\vspace{2mm}
It seems likely that pure particle weights
of physical interest are regular. But we have not been able to
establish this property in general.

\vspace{2mm}
\nin {\bf Proposition 3.4:}  {\em  Let
$\ketl{\cdot}$
be a pure particle weight. There exists a unique vector
$p_{\la} \in {\RR}^d \backslash \overline{V}_-$
such that the continuous unitary representation
$x \rightarrow U_{\la}(x)$
of space-time translations given by
$$ U_{\la}(x) \ketl{L} \doteq e^{i p_{\la} x} \ketl{ \ax (L)}, \
L \in \Li $$
has Lorentz invariant spectrum in
$\overline{V}_+$.
If $\ketl{\cdot}$ is regular, then
$p_{\la} \in \overline{V}_+ \backslash \{ 0 \}$.
}

\vspace{2mm}
The interpretation of this result is obvious: pure particle weights
$\ketl{\cdot}$
have sharp energy-momentum
$p_{\la}$.
Appealing to the picture of a plane wave, the role of the operators
$L \in \Li$
becomes clear then: they localize the wave and thereby produce a
normalizable state. Incidentally, this explains the terminology
"left ideal of localizing operators". It is not excluded by our
results that
$p_{\la}$
is spacelike. The corresponding "tachyonic" particle weights
would be non-regular, but become positive energy states after
localization.

Let us next discuss the behaviour of particle weights under
Lo\-rentz-trans\-for\-ma\-tions. Given a pure particle weight
$\ketl{\cdot}$
and any Lorentz transformation
$\Lambda$,
we can define another particle weight
${\ket{\cdot}}_{\la, \Lambda}$,
setting
\be  {\ket{\cdot}}_{\la, \Lambda} \doteq  \ketl{{\aL}^{\! -1} (L)},
\ L \in \Li. \ee
One easily verifies that
${\ket{\cdot}}_{\la, \Lambda}$
has energy momentum
$\Lambda p_{\la}$.
Restricting
$\Lambda$
to the little (stability) group
${\cal R} (p_{\la})$ of $p_{\la}$
one can thus explore the internal ("spin") degrees of freedom of
$\ketl{\cdot}$.
Here we restrict attention to the physically relevant case, where
the number of these degrees of freedom is finite (finite particle
multiplets).

\vspace{2mm}
\nin {\em Definition:} A pure particle weight
$\ketl{\cdot}$
is said to have finite spin if the space of sesquilinear forms on
$\Li \odot \Li$
generated by
$ {}_{\la , \Lambda} {\bracket{\cdot}{\cdot}} {}_{\la, \Lambda},
\ \Lambda \in {\cal R} (p_{\la})$
is finite dimensional.

\vspace{2mm}
\nin {\bf Proposition 3.5:}  {\em Let
$\ketl{\cdot}$
be a pure particle weight with finite spin and let
$( {\pi}_{\la}, {\cal H}_{\la} )$
be the corresponding GNS-representation of
$\Al$.
On
${\cal H}_{\la}$
there exists a continuous unitary projective representation
$\Lambda \rightarrow V_{\la} ( \Lambda )$
of the little group
${\cal R} ( p_{\la} )$
such that
$$ V_{\la}( \Lambda ) {\pi}_{\la} (A) V_{\la} (\Lambda)^{-1}
= {\pi}_{\la} ( \aL (A) ), \: A \in \Al. $$
Moreover, the linear space of particle weights generated by
$V_{\la} ( \Lambda )^{-1} {\ket{\cdot}}_{\la, \Lambda},
\Lambda \in {\cal R}(p_{\la} ) $ is finite dimensional. }

\nin {\em Remark:} We recall that there may not exist
a representation of the full Lorentz group (infraparticle problem).
\vspace{2mm}

Let us discuss the consequences of this result for regular particle
weights in physical space-time $d=4$. There we have to distinguish
two cases: if the mass-square
$p_{\la}^2$
is positive, the little group of
$p_{\la}$
is isomorphic to O(3). Hence the projective representation
$V_{\la}$
can be lifted to a true representation of its covering group SU(2),
and the particle weights can be decomposed into components of
fixed (half) integer spin. So in this case we arrive at the same
conclusions as Wigner. If
$p_{\la}^2 = 0$
the situation is different, however. In this case the little group
of
$p_{\la}$
is isomorphic to E(2), and it may well be that
$V_{\la}$
cannot be lifted to a true representation of its two-fold covering,
if Lorentz boosts are spontaneously broken in the representation
$( {\pi}_{\la}, {\cal H}_{\la}) $.
The spin (helicity) of particle weights can still be defined in these
cases, but it does not have to be (half) integer. Thinking of the
spin-statistics theorem, this leaves open the possibility of massless
infraparticles in physical space-time with unusual statistics.

There are further results which substantiate the heuristic expectation
that particle weights describe the timelike asymptotic particle
content of states (such as specific localization properties of the
systems described by particle weights, local normality properties
etc. \cite{BP}). But the results quoted so far may suffice for the
motivation of the following.

\vspace{2mm}
\nin {\em Postulate:} Particles appearing at asymptotic times in few
body systems are described by (pure) particle weights
$\ket{\cdot}$,
mapping the left ideal
$\Li$
of localizing operators into some Hilbert space
${\cal H}$.
The ensembles which can experimentally be prepared by the process
of localizing particles are described by the range of
$\ket{\cdot}$.

\vspace{2mm}
What have we gained by this point of view? On one hand a formalism
which allows a unified treatment of particles, infraparticles,
tachyons (should they exist), etc. Having solved this conceptual
problem, it amounts to a well defined mathematical problem to
determine the specific properties of these systems
(Poincar\'e covariance versus spontaneous breaking of Lorentz boosts,
negative mass squares, etc.).
On the other hand, our point of view is the basis for a simple
and general formula by which the asymptotic particle content
can be extracted from any state
$\omega$,
\be \bracket{L}{L} = \mbox{"} \lim_{t} \mbox{"}
\int \dsx \, \omega ( {\alpha}_{t, \sx}( L^* L ) ),
\: L \in \Li. \ee
(Here the quotation marks indicate the possible need for time
averages and the transition to subnets.) It is a striking fact,
first stated implicitly in \cite{AH} in a more restricted setting,
that in this way the (charged) particle content of a theory can be
uncovered by using only the local observables in the vacuum sector.

We conclude this discussion with a list of open problems and
remarks.

i) When does a theory have a non-trivial asymptotic particle
structure, i.e. under which conditions are the limits
$\bracket{\cdot}{\cdot}$
in (13) non-zero? It would be of great interest to formulate these
conditions in general terms (using the underlying net structure)
since it would clarify the question of how the particle aspects
are encoded in the local properties of a theory. It seems likely
that the following two conditions are the key ingredients: the
condition of primitive causality \cite{HS}, being the abstract
expression of the existence of a dynamical law, and a condition
of causal independence \cite{BWi}, now called split-property
\cite{DL}. A heuristic argument to this effect has been given in
\cite{B4}, but it has not yet matured into a proof.

ii) When do the asymptotic particle configurations stabilize, i.e.
under which conditions can the quotation marks in (13) be removed
(existence of limits)? As is clear from quantum mechanical
scattering theory \cite{AC}, a positive answer will depend on
decent "phase space properties" of the theory. In the present
setting, this requirement can be expressed in terms of compactness
or nuclearity conditions (cf. \cite{BP2} for a comprehensive account
and references).

iii) Is there a general collision theory based on particle weights?
Some interesting progress on this problem has been made in
collaboration with U. Stein. We have developed a novel algorithm which
seems suitable for the construction of collision states of (infra-)
particles of prescribed charge, spin, mass and momentum (possibly
accompanied by an unspecified cloud of massless particles with
arbitrarily small total energy), cf. \cite{BPS}. It is based on
the algebra
$\Co = {\Li}^* \Li$
and does not require the knowledge of charge carrying fields. It
thereby fills the remaining gap in \cite{AH} in the direct
construction of collision cross sections from observables in the
vacuum sector. A proof that the method works has so far
only been given in massive, asymptotically complete theories \cite{S}.
But the method can be directly applied (and tested) in any model.

iv) What is the statistics of particle weights? An approach to this
problem could be based on an extension of the DHR-analysis
\cite{DHR1+2} to superselection sectors describing non-localizable
charges (such as the electric charge), and there are some promising
ideas to this effect \cite{BDMRS}.

Another approach could look as follows: it seems possible to construct
with the help of the operators
$\Li \in \Li$
loops in the space of
collision states, describing the exchange of two (or more) particles
of the same kind. In order to determine the statistics of the
particles one has to lift these loops to the space of collision
state vectors. The appropriate lift seems to be the Berry lift,
respectively its generalizations (cf. \cite{U} and references quoted
there), and the statistics would then be given by the
corresponding monodromy operator (Berry phase). Some preliminary
calculations indicate that this definition of statistics
gives, in the case of localizable charges, the same answers as the
definition of statistics in \cite{DHR1+2}. Another question,
closely related to this problem is: When do there exist antiparticle
weights?

v) When does a theory have a complete particle (weight)
interpretation? A definite mathematical formulation of this question
has not yet been accomplished. (Note that an unambiguous scattering
matrix does not exist in the case of infraparticles.) But there is
some closely related problem which is of interest in its own right:
in high energy collision experiments only some very
limited aspect of completeness is really tested, which primarily
relies on the existence of conservation laws. One prepares, for
example, (initial) states
$\omega$
with relatively sharp energy momentum and checks whether the
energy-momentum carried by the outgoing particles adds up to the same
result. So there are two quantities involved: the energy-momentum
pumped into the state
$\omega$
in the process of its preparation, which may be identified with
$\omega (P)$,
$P$ being the energy-momentum operator. And the energy-momentum
carried by the asymptotic particles, which according to
Proposition 3.4 can be identified with the values
$p_{\la}$
corresponding to the particle weights
$\ketl{\cdot}$
appearing in the asymptotic decomposition (11) of
$\ket{\cdot}$.
Hence, in accord with the experimental facts, there should
hold
\be \omega (P) = \int d \mu ( \la ) \, p_{\la}, \ee
where $\mu$ is the measure appearing in the decomposition
(11),\footnote{An
unambiguous definition of $\mu$ requires a normalization of the
particle weights $\ketl{\cdot}$, a problem not touched upon here}
and a similar
relation should hold for the other conserved quantities (charge,
spin, etc.) which can be attributed to particles. It seems
that such a weak form of asymptotic completeness suffices for a
consistent interpretation of most collision experiments. We
conjecture that relation (14) holds in all theories admitting a
local stress energy tensor.
\section{Ultraparticles}

First some word about terminology: the term infraparticle was
coined by B. Schroer \cite{Sch} as a catch phrase alluding to the
infrared problems which arise in models with long range forces.
In the preceding section we have seen how infraparticles (like
particles) appear at large distances in states of moderate energy.
Thus in this respect the term infraparticle also fits well. In the
present section we are interested in the particles appearing at short
distances in the limt of arbitrarily high energies. We call these
particles {\em ultraparticles}. Some of them may be infraparticles
as well, e.g. leptons, others do not show up at large scales,
e.g. partons. The latter phenomenon (confinement) may be
understood as an infrared problem. But the term ultraparticle
seems more appropriate also in these cases since it specifies
the domain where these particles are observed.

Within the perturbative treatment of field theory, a powerful tool
for the analysis of the ultraviolet properties of models is based
on scaling transformations of fields (renormalization group
transformations \cite{RG}). They allow an interpretation of the
theory at small scales and have led to fundamental concepts such
as the notion of asymptotic freedom. Although well known, let us
recall that the scaled fields are regarded as elements of another
theory at the original scale with e.g. different (running)
coupling constants. Phrased differently: the interpretation of the
original theory at small scales is accomplished in terms of other
theories, but at fixed scale. We emphasize this fact in order to
make clear from the outset that the particle structures which are
of interest here will (and can) not be described by states or
weights on the given algebra of observables
$\Al$.

The utility of scaling transformations for the structural analysis
in algebraic quantum field theory has first been realized by
Roberts \cite{R} in the special case of scale-invariant theories.
The methods were later refined in \cite{HNS} (cf. also \cite{F}
and \cite{FH}) to cover also the case of non-scale-invariant
theories. These proposals contain, however, a conceptually
unsatisfactory element since they depend on the
existence of pointlike
quantum fields. From the algebraic point of view such fields (if
they exist) are usually interpreted as a kind of coordinatization
of the local algebras. This raises then the questions: does the
"principle of local stability", proposed in \cite{HNS}, depend
on a specific choice of such coordinates? Do the results on the
structure of local algebras in \cite{F} depend on the existence
of such coordinates?
Moreover, such interesting questions as
to when there exist quantum fields describing the ultraviolet
properties of a net of observables, or what is the possible ultraviolet
structure if there are no such fields are put aside from
the outset.

In joint work with R. Verch, we have set up a framework which
allows the systematic discussion of these problems \cite{BV}, and
we are presently investigating the ultraviolet properties of local
nets with emphasis on the ultraparticle aspects. In the
following an outline of the basic ideas in our approach is given,
and some pertinent results with regard to the ultraparticle
problem are discussed.

What is required in order to express the ideas of the
renormalization group in the algebraic setting? One first needs a
method to proceed from the local algebras
$\AO$
at scale
$\la = 1$,
say, to algebras
$\AlO$
describing the observations made at smaller scales
$\la > 0$
in the regions
$\la \Oo$,
i.e.
$\AlO = {\cal A} ( \la {\cal O} )$.\footnote{Note that the map
$\Oo \rightarrow \AlO $
defines again a local net over Minkowski space with
Poincar\'e-transformations given by
${\alpha}^{(\la)}_{x} \doteq  {\alpha}_{\la x}$
and
${\alpha}^{(\la)}_{\Lambda}  \doteq {\alpha}_{\Lambda} $}
There are many ways of introducing such maps: conditional
expectations, canonical endomorphisms based on the
Tomita-Takesaki theory, scaling transformations of underlying
quantum fields, etc. Apart from an important physical constraint
which will be discussed below, it should not really matter which
map one takes since they yield the same nets (they merely
reshuffle the operators in the local algebras in different ways).
So why not consider them all?

A convenient way of expressing this idea is provided by the concept
of a scaling algebra
$\Alu$
associated with any given local net
$\Al$.
The elements $ \Au $ of
$\Alu$
are bounded functions of the scaling parameter
$\la \in ( 0, \infty )$
with values in
$\Al$,
i.e.
$\Aul \in \Al, 0 < \la < \infty$.
The algebraic operations in
$\Alu$
are pointwise defined by the corresponding operations in
$\Al$,
and there is a C$^*$-norm on
$\Alu$
given by
\be || \Au || = \sup_{\la} || \Aul ||. \ee
The local structure of
$\Al$
is lifted to
$\Alu$
by setting
\be \AOu = \{ \Au \in \Alu : \Aul \in \Al ( \la \Oo ), 0 < \la < \infty
\}. \ee
Hence
$\Oo \rightarrow \AOu $
defines a net over Minkowski space, and it suffices for our purposes
to assume that
$\Alu$ is the C$^*$-inductive limit of the local algebras
$\AOu$.
It is easily verified that the net is local, i.e.
$\AOiu{1} \subset {\AOiu{2}}'$ if $\Oi{1} \subset {\Oi{2}}'$.
Moreover, the action of the Poincar\'e group on
$\Al$
can be lifted to a covariant action on the net
$\Alu$,
induced by automorphisms
$\axu , \aLu \in \mbox{Aut} \Alu$
given by
\be ( \axu ( \Au ) )_\la \doteq {\alpha}_{\la x} ( \Aul ), \, \,
( \aLu ( \Au ) )_\la \doteq \aL ( \Aul ). \ee
Not quite unexpectedly, there appears an additional symmetry in this
setting: dilations. This group acts by automorphisms
$\dmu , 0 < \mu < \infty,$
on the net, given by
\be ( \dmu ( \Au ) )_\la \doteq \Au_{\mu \lambda}. \ee
This formalism nicely expresses the geometrical aspects of scaling
transformations. But it is probably of little use as it stands
since there is some important physical constraint missing: we want to
change the length scale, but the other physical units should stay
constant. The velocity of light, $c=1$, has not been changed by our
procedure since we have scaled space and time by the same factor
$\la$.
But what about Planck's constant? Many of the (only abstractly
characterized) functions $\Au $ will describe a situation, where
simultaneously with length also Planck's constant is scaled in some
uncontrolled manner. Thus we are  not yet in a setting
corresponding to the framework of the renormalization group, unless
we manage to get rid of these unwanted elements.

In a first attempt to solve this problem, we have tried to isolate in
$\Alu$
some "Planck-ideal" which, by taking quotients, fixes the value
of Planck's constant at each scale. But this idea did not work. The
final solution
of the problem turned out to be amazingly simple and consists
of a direct characterization of the quotient of $\Alu$, based on the
follwing
idea: since the product of the dimensions of length and momentum
is equal to the dimension of action, we have to scale momentum
by
$\la^{-1}$
if we scale length by
$\la$
and want to keep Planck's constant fixed. The energy-momentum scale
can be set by determining the energy-momentum transferred by
"standard operators" to physical states. Hence if
${\underline{A}}_1 \in \Al$
is such an operator which at scale
$\la = 1$
transfers energy momentum
$\Gamma$,
where
$\Gamma$
is some compact subset of ${\RR}^d$, then its counterpart
$\Aul \in {\cal A}_{\la}$
at scale
$\la$
should transfer energy-momentum
$\la^{-1} \Gamma$.
Now the local operators
$A \in \Al$
can transfer any energy-momentum. But they are "quasi-local in
momentum space" as a consequence of their continuity properties
with respect to space-time translations. In fact, given
$A \in \Al$
and
$\varepsilon > 0$
one can find some operator
$A_{\Gamma}$
with compact energy-momentum transfer
$\Gamma$
such that
$|| A - A_{\Gamma} || < \varepsilon$.

After these remarks it is straightforward now to express the
requirement that Planck's constant is kept fixed under scaling
transformations: we have to restrict attention to functions
$\Au$ which, at every scale
$\la$,
can be approximated with uniform accuracy
$\varepsilon > 0$
by operators with energy-momentum transfer
$\la^{-1} \Gamma$,
where
$\Gamma$
is some fixed compact set. Phrased in formula form:
$\sup_\la || \Aul - {\Au}_{\la, \la^{-1} \Gamma } || < \varepsilon $.
After a moments reflection one sees that this condition is
equivalent to the requirement that
$\sup_\la || {\alpha}_{\la x} ( \Aul ) - \Aul || \rightarrow 0 $
as
$x \rightarrow 0$,
which simply means that the function
$x \rightarrow \axu ( \Au ) $
is to be strongly continuous. Thus the condition
"$\hbar = 1$
at all scales" amounts, in the framework of the scaling algebra
$\Alu$,
to the condition of strong continuity under space-time
translations.

The same considerations, as indicated above for the case of
energy-momen\-tum, apply of course also to angular momentum and
lead to the requirement that also the functions
$\Lambda \rightarrow \aLu ( \Au ) $
have to be strongly continuous. Hence we are led to the

\vspace{2mm}
\nin {\em Definition:} Let
$\Al$
be any local net complying with the assumptions given in Sect. 2.
The canonically associated scaling algebra
$\Alu$ is the C$^*$-inductive limit of the local net
$\Oo \rightarrow \AOu $
of functions $ \Au$  which are strongly continuous with respect to
the action of the Poincar\'e group, cf. relations (16) and (17). \\
(We note that the concept of scaling algebra can also be applied to
local nets on curved space-time \cite{BV}.)

\vspace{2mm}
\nin {\em Remark:} Within this setting the significance of the
automorphisms
$\dmu , 0 < \mu < \infty $
becomes clear: they induce scaling (renormalization group)
transformations on
$\Alu$.
In view of this fact it may seem natural to assume that $ \dmu $
acts strongly continuously on
$\Alu$
(which amounts to further restrictions on the elements of
$\Alu$).
It would, however, be too much to require continuity of the
functions
$\mu \rightarrow \dmu ( \Au ) $
at
$\mu = 0$,
as this would imply that
$\mbox{n-} \! \lim_{\mu \rightarrow 0} \pi ( {\Au}_{\mu} ) = const
\cdot 1 $
in all positive energy representations
$\pi$
of
$\Al$.
The corresponding functions $\Au$ would therefore not be suitable
to explore the ultraviolet properties of a theory. In the following
discussion we do not anticipate any continuity properties of
the scaling transformations.

We next discuss how the structure of the states
$\omega$
at small scales can be analyzed with the help of the scaling algebra.
To this end we have to lift these states to
$\Alu$.

\vspace{2mm}
\nin {\em Definition:} Let
$\omega$
be any state on
$\Al$.
Its canonical lift
$\ou$
to the scaling algebra
$\Alu$
is given by
\be \ou ( \Au ) \doteq \omega ( {\Au}_{\lambda = 1} ), \, \Au \in \Alu.
\ee

Before we proceed let us clarify the physical interpretation
of the states
$\ou$
on
$\Alu$.
In order to recover this interpretation, one has to regard
$\ou$
as a (projected) state on the quotient
$\Alu / ker \, \pu$,
where
$\pu$
is the GNS-representation induced by
$\ou$ and $ ker \, \pu$ the kernel of $\pu$.
To see this, note that if
$\pi$
is the GNS-representation induced by
$\omega$,
the nets
$\Alu / ker \, \pu$
and
$\Al / ker \, \pi$
($= \Al$
if
$\pi$
is faithful) are isomorphic\footnote{Two nets are said to be
isomorphic if there exists an isomorphism between the respective
global algebras of observables, which maps the local algebras
corresponding to the same regions
$\Oo$
onto each other. Two such nets describe "the same physics"}, an
isomorphism $\phi$ being given by
$\phi ( \pu ( \Au ) ) = \pi ( {\Au}_{\la = 1} ), \Au \in \Alu$.
This isomorphism also intertwines the projected actions of
the Poincar\'e group on the respective nets if the kernel of
$\pi$,
and hence that of
$\pu$,
is Poincar\'e invariant. These properties characterize
$\phi$
up to internal symmetries. Moreover, if
$proj \,  \ou$
denotes the state obtained by projecting $\ou$ to the quotient
$\Alu / ker \, \pu$,
there holds
$proj \, \ou \circ  {\phi}^{-1} = \omega $.
Thus the state
$proj \, \ou $
on the net
$\Alu / ker \, \pu$
contains the same physical information as
$\omega$,
provided there exist no internal symmetries of the algebra of
observables
$\Al$.
If there are such symmetries, one still knows that
$proj \, \ou $
describes some state on
$\Al$
which lies on the orbit of
$\omega$
under the action of the internal symmetry group. Hence if
$\omega$
is a vacuum state, then
$proj \, \ou $
also describes a vacuum state which coincides with
$\omega$,
unless the internal symmetry group is spontaneously broken. This
information suffices for our purposes.

In order to analyze the state
$\omega$
at scale
$\la$,
we have to apply the scaling transformation
${\underline{\delta}}_{\lambda}$
to its canonical lift
$\ou$,
giving the state
$\ou \circ {\underline{\delta}}_{\lambda}$
on
$\Alu$.
If we proceed as above and consider the projected state
$proj \, \ou \circ {\underline{\delta}}_{\lambda}$
on the net
$\Alu / ker \, {\pu}_{\la}$,
where
${\pu}_{\la}$
is the GNS-representation induced by
$\ou$,
we recover the state
$\omega$
on the net
${\Al}_{\la}$,
up to internal symmetry transformations. (The internal symmetry group
is of course the same at each scale
$\la > 0$.)

We are here interested in the structure of the physical states
$\omega$
at scales
$\la \rightarrow 0$,
so we have to consider the corresponding sequence of states
$\ou \circ {\underline{\delta}}_{\la}$
in this limit. Here some technical nuisance appears since these
sequences do not converge. To explain the simple origin of this
phenomenon, let us consider the following toy model: we take as our
algebra of observables the complex numbers,
$\Al = \CC$.
The corresponding scaling algebra
$\Alu$
is then the algebra of all bounded $c$-number functions
$\Au : \, ( 0 , \infty ) \rightarrow \CC $
with norm
$|| \Au || = \sup_{\la} | \Aul | < \infty $.
The canonical lift of the (unique) state
$\omega$
on
$\Al$
to the scaling algebra
$\Alu$
is given by
$\ou ( \Au ) = {\Au}_{1}$,
and
$\ou ( {\underline{\delta}}_{\la} ( \Au ) ) = \Aul $.

Since we may not assume from the outset that all functions $\Au$
converge for
$\la \rightarrow 0$
(cf. the previous remark) it is clear that the sequence
$\ou \circ {\underline{\delta}}_{\la} $
will not converge in this limit. The problem is that the scaling
algebra provides us with an abundance of mappings from the algebra
$\Al$
at scale $\la = 1$
to the corresponding algebras
$\Al_\la$  $\mbox{(} = \Al \mbox{)}$
at other scales
$\la$.
(This was, after all, our initial idea.) To proceed,
we make use of the fact that the set of states
$\ou \circ {\underline{\delta}}_{\la} $
is relatively compact in the weak *-topology induced by
$\Alu$.
Thus there exist (many) convergent subnets for
$\la \rightarrow 0$.
In a sense, each convergent subnet corresponds to some particular
choice of the above mentioned mappings.

Let us denote the limits of these subnets by
$\oziu , \iota \in \II$
($\II$ being some index set). Recalling the physical interpretation
of states
$\omega$
at scale
$\la$
in terms of
$\ou \circ {\underline{\delta}}_{\la} $,
we consider the projected states
$proj \oziu$
on
$\Alu / ker {\pu}_{0  , \iota}$,
${\pu}_{0  , \iota}$
being the GNS-representation induced by
$\oziu$.
It is evident that in our model
$\Alu / ker {\pu}_{0  , \iota}$
is isomorphic to
$\Al$.
Thus the abundance of functionals
$\oziu , \iota \in \II$
on
$\Alu$
describes a single "physical state", the state
$\omega$
from which we started. (This is of course no surprise.)

After these preliminary remarks the reader will guess how we proceed:
given a state
$\omega$
on
$\Al$,
we consider the corresponding family of states
$\ou \circ {\underline{\delta}}_{\la} $
on
$\Alu$ and its weak *-limit points for $\la \rightarrow 0$:
$\oziu , \iota \in \II$.
We then propose as

\vspace{2mm}
\nin {\em Working hypothesis:} Let
$\omega$
be a state on
$\Al$.
The limit points
$\oziu , \iota \in \II$
of
$\ou \circ {\underline{\delta}}_{\la} $
for
$\la \rightarrow 0$,
projected on the respective nets
$\Alu / ker {\pu}_{0  , \iota}$,
(${\pu}_{0 , \iota}$
being the GNS representation induced by
$\oziu$),
describe the structure of the state
$\omega$
in the scaling limit
$\la \rightarrow 0$
(possibly modulo internal symmetry transformations).

\vspace{2mm}
If the nets
$\Alu / ker {\pu}_{0  , \iota}$
in this statement are all isomorphic, one can interpret the
asymptotic structure of
$\omega$
in terms of a single theory. We then say that the original theory
has a unique scaling limit. (A justification for this terminology
will be given below.) But
this does not mean that the states
$proj \oziu$
all coincide, there may appear new "degrees of freedom" in the
scaling limit. The second possibility is that not all of the nets
$\Alu / ker {\pu}_{0  , \iota}$
are isomorphic. It means that the theory has no fixed interpretation
in the scaling limit, it looks different at different scales.

After this lengthy preparation of our tools, let us turn now to
the actual analysis of the scaling limits. To this end we first
have to fix the states which we want to consider. Being interested
in states in high-energy physics, we pick the set
${\cal S}_0$
of all states which are locally normal \cite{HKK} with respect to
a given pure Poincar\'e invariant vacuum state
$\oz$ on $\Al$.
This set of states is actually quite big, it does not only contain all
finite energy excitations of
$\oz$,
but in general also thermal states, etc. Our first result says that it
does not matter which state
$\omega \in {\cal S}_0$
we consider, they all describe the same structure in the scaling
limit. In fact there holds a stronger result.

\vspace{2mm}
\nin {\bf Lemma 4.1:} {\em Let
${\omega}_1 , {\omega}_2 \in {\cal S}_0$.
Then
$$ \lim_{\la \rightarrow 0} ( {\ou}_1 \circ {\underline{\delta}}_\la
- {\ou}_2  \circ {\underline{\delta}}_\la ) ( \Au ) = 0, \,
\Au \in \Alu.$$  }

In simple terms: all states
$\omega \in {\cal S}_0$
look alike at small scales. (This justifies our terminology
concerning the uniqueness of the scaling limit of a theory.) But
how do they look like? Here is the answer for
$d > 2$
dimensions:

\vspace{2mm}
\nin {\bf Proposition 4.2:} { \em Let
$\omega \in {\cal S}_0$.
Then each weak *-limit point
$\oziu$
of
$\ou \circ {\underline{\delta}}_{\la}$, $\la \rightarrow 0$,
is a pure Poincar\'e invariant vacuum state on
$\Alu$.
(Thus it is also a pure, Poincar\'e invariant vacuum state if
projected to the net
$\Alu / ker {\pu}_{0  , \iota}$,
since the kernel of the GNS-representation
${\pu}_{0  , \iota}$,
induced by
$\oziu$,
is Poincar\'e invariant.) }

\vspace{2mm}
This result is the key to a general understanding of the scaling
limits of quantum field theories. Let us first indicate how it can be
used for their classification. There are the following three
possibilities.

1. The theory has a {\em classical} scaling limit, i.e.
$\Alu / ker {\pu}_{ 0 , \iota} \simeq \CC$
for
$\iota \in \II$.
A trivial example of this kind is obtained if one (re)defines
the algebras
$\AO \subset \Al$
to be equal to
$\CC \cdot 1$
for small regions
$\Oo$.
This artificial possibility is not what we here have in mind
(and it could be excluded by a condition of additivity on the net
$\Al$).
But our structural result seems to indicate that there may well
exist theories which become classical in the scaling limit.
Examples of this kind could be some of the models envisaged by
J. Klauder, cf. these Proceedings. We note that under these
circumstances the "classical ideal"
in $\Alu$
(generated by all commutators) is annihilated in the scaling limit,
i.e. it is a proper ideal of
$\Alu$.

2. The theory has a unique {\em quantum} scaling limit, i.e. the
(isomorphic) nets
$\Alu / ker {\pu}_{0  , \iota} , \iota \in \II$
are non-abelian. This case corresponds to theories which have
(in the language of the theory of the renormalization group) an
ultraviolet fixed point. Of particular interest are the
asymptotically free theories. According to the folklore, in such
theories the underlying fields have almost canonical dimensions (apart
from logarithmic corrections) and become free massless fields in
the scaling limit. We therefore propose to characterize such
theories in the algebraic setting by the condition that (for
even dimensions $d$) the "Huygens-ideal" in $\Alu$
(generated by all commutators of pairs of operators
which are localized in strictly timelike separated regions)
is annihilated in this limit.

3. The theory has no scaling limit, i.e. not all of the
nets
$\Alu / ker {\pu}_{0  , \iota} , \iota \in \II$
are isomorphic.\footnote{It may very well be that "physics"
behaves like this scenario. Then the hope that one may guess a "grand
unified theory" which describes physics at all scales would probably
be an illusion. But this is not a real problem. After all, high
energy experiments can only be carried out up to the scale
$\la_{budget}$,
and it should therefore be possible to understand the prevailing data
in terms of the simpler cases 2 or 1}

We mention as an aside that, as a consequence of Proposition 4.2 and
a result in \cite{F}, the weak closures of the local algebras $\AO$
(with respect to the folium of states ${\cal S}_0$) are of
type $\mbox{III}_1$ for all double cones $\Oo$, if the underlying theory
does not have a classical scaling limit.

Let us turn now to the discussion of ultraparticles, the actual
subject of this section. So where are they? Proposition 4.2 provides
the tools to answer also this question: the ultraparticles are the
particles (in the sense of the previous section) of the theory
in the scaling limit. This
characterization is meaningful, since in the scaling limit we are
still in the framework of positive energy (vacuum) representations
of some local net, and the methods of the preceding section can be
applied. But even more important: this characterization complies
with the physical picture which led to the successful
interpretation of experimental data. We emphasize that in our analysis
the "ultraparticles" are not put in by hand, they appear naturally
as a possible structure within the general framework of local
quantum field theory. This fact leads us to propose as a

\vspace{2mm}
\nin {\em Postulate:} The ultraparticle content of all states
in a theory which
look locally like few body systems (states of finite energy)
is described in terms of the particle content of the vacuum
representations of the scaling limit theory, cf. Proposition 4.2 and
Sect.3.

\vspace{2mm}
We refrain from giving here a survey of our partial results on the
structure of ultraparticles. But we would like to point
out one interesting aspect of our analysis: although we started
from local (gauge-invariant) observables, we expect to recover in
the scaling limit the full gauge-group of the underlying theory,
including the {\em local gauge group}. The structure of the full gauge
group is expected to be encoded in the global gauge group of the
scaling limit theory since
there is no difference between global and local symmetry
transformations at a fixed space-time point. Since the global
gauge group of any theory can be reconstructed from the net of local
observables in a systematic manner \cite{DR}, we have thus found
a promising approach to identify gauge quantum field theories, starting
from local observables.

We conclude this section with an extract from the long list of
intriguing question in this new setting of algebraic quantum field
theory.

i) What is the ultraparticle content of a given {\em state}? So far we
have only described the ultraparticle content of a theory (by
looking at the particle content of all states in the vacuum sector
of the scaling limit theory). For the analysis of the specific
ultraparticle structure of a given state
$\omega$
one can probably proceed as in Sect. 3. One first distinguishes
some left ideal
$\widetilde{\underline{\Li}} \subset \Alu$
which annihilates all vacuum states on
$\Alu$.
But now the operators in
$\widetilde{\underline{\Li}}$
have to localize the ultraparticles in momentum space (they have to
cut off large momenta). In complete analogy to the discussion in
Sect. 3 one could then try to extract the ultraparticle content of
$\omega$
from the leading contributions in
$\ou \circ {\underline{\delta}}_{\la}
( \widetilde{\underline{L}}^*
\widetilde{\underline{L}} ) $,
$\widetilde{\underline{L}} \in {\widetilde{\underline{\Li}} }$,
for small
$\la$.
A further step would be the formulation of a collision theory for
ultraparticles.

ii) When is a theory asymptotically free? We have characterized
asymptotically free theories in an ad hoc way by adopting the
folklore and converting its algebraic aspects into a condition which
fits naturally into our framework. But it would be desirable to
establish this condition in an intrinsic way. A possible approach
(in massive theories like QCD) could be based on a study of the
assumption that the S-matrices
$S_{\la}$
of the theory at scale
$\la$
tend to unity in the scaling limit
$\la \rightarrow 0$.

iii) What is the field content of a theory? The scaling algebra
seems to be the adequate instrument to analyze also this question
and to explore for example the interesting ideas outlined in
\cite{H1}.

iv) What is the structure of the theory at large scales
$\la \rightarrow \infty$?
The general concept of scaling algebra may also be relevant for a
study of other "scaling limits" such as $c \rightarrow \infty$
or $\hbar \rightarrow 0$.

\section{Thermoparticles}

Let us turn now to the analysis of the particle-like constituents
of thermal equilibrium states. As will become clear, these entities
have to be distinguished from quasiparticles, describing the
collective excitations of thermal states, and also from the other
particle structures analyzed so far. We therefore propose to call
them {\em thermoparticles}.\footnote{Our terminology does not
intend to suggest that for a description of these particles one has
to rely on the framework of "thermofield theory"}

How does one determine the thermoparticle content of a given
KMS-state
$\omega$
on
$\Al$?
A complete answer to this question is not yet known, but for the
special case of spatially homogenous KMS-states some promising
results have been obtained in an investigation with J. Bros
\cite{BB1}, cf. also \cite{BB2}.
In this approach we proceed again
from the heuristic idea that thermoparticles, like other particles,
are almost point-like objects. It should therefore be meaningful
to consider the situation, where such a particle is placed in an
equilibrium state into some region, about a space-time point
$x$,
whose size is small compared to the mean volume occupied by the
other constituents. (Phrased differently: the de Broglie wavelength of
the particle should be small compared to its mean free path.) This
particle will either collide with constituents of the state
and thereby disappear in the thermal background. Or it will manage
to propagate to another space-time point $y$ without intermediate
collisions. The latter events should give rise to particle-like
correlations between operations performed in almost point-like
regions due to the creation and annihilation of a single particle,
respectively hole. The probability for such events may be small
because of the effects of the thermal background. Nevertheless, the
correlation functions of point-like localized operators may be
expected to contain the desired information about thermoparticles
in a clear-cut form.

Let us translate now these heuristic ideas into our mathematical
setting. In order to simplify the discussion we assume, for the
time being, that the constituents of the underlying equilibrium
state
$\omega$
are "neutral" in the sense that they can be created by local
operations described by certain specific operators
$A \in \Al$.
{}From the preceding discussion it is clear that most operators in
$\Al$
are not suitable for our purposes, since they describe
operations in finite space-time regions
$\Oo$
which in general will contain many constituents. The resulting
excitations of
$\omega$
may have a quasiparticle structure, but it seems hopeless to
extract from them some relevant information about the thermoparticles.
Since we do not know from the outset how small
$\Oo$
has to be chosen in order to avoid this problem, it seems appropriate
to proceed to the idealization, where the localization region of the
underlying operators
$A \in \Al$
is contracted to a point. Thus, as in the preceding section, we are
led to consider sequences of local operators
$A_\la \in {\Al} ( \la \Oo )$,
where
$\la$
tends to $0$. But now we want to use these operators to create
particle-like excitations of the equilibrium state which can be
observed at the original scale
$\la = 1$.

As explained before, we expect that information about the
thermoparticles can be extracted from the correlation functions of
the operators
$A_\la$.
Assuming that
$\omega$
is invariant under space-time translations, these functions can be
represented in the form
\be \omega ( {\alpha}_y ( A_\la )^* \ax ( A_\la ) ) =
\int d {\mu}_\la ( p ) \, e^{ip(x-y)}, \ee
where
${\mu}_\la$
is some positive measure. The dominant contribution to this measure
will in general be a discrete part at $p=0$, corresponding to those
events where the operation described by
$A_\la$
has no effect on the state
$\omega$.
This uninteresting part has to be removed from
${\mu}_\la$
(e.g. by subtracting from
$A_\la$
its expectation value
$\omega ( A_\la )$
if
$\omega$
describes a pure phase), and we assume in the following that this
has been done. The remaining part of the measure describes the
correlations between well-localized excitations of
$\omega$.

In order to extract from
${\mu}_\la$
the contributions due to the propagation of a single thermoparticle,
we have to proceed to the limit
$\la \rightarrow 0$.
Here some further preparations are necessary. On one hand, the
energy-momentum content of the excitations produced by
$A_\la$
will become infinite in the limit
$\la \rightarrow 0$.
As in the case of quantum fields, this effect can be controlled
by taking suitable space-time averages of the operators
$\ax ( A_\la )$,
i.e. the respective limit has to be understood in the sense of
distributions. On the other hand, the probability that the operators
$A_\la$
create some excitation of
$\omega$
will tend to $0$ for
$\la \rightarrow 0$
if the norms of these operators are uniformly bounded. In order to
compensate the latter effect, the measure
${\mu}_\la$
has to be renormalized. This is accomplished by picking any
$ N \in \NN$
and fixing a corresponding normalization constant
\be Z_\la = \int d {\mu}_\la ( p )  ( 1 + | p | )^{- N} , \, \,
\la > 0. \ee
The resulting renormalized correlation functions
\be W_\la ( x ) = {Z_\la}^{-1} \int d {\mu}_\la ( p ) \,
e^{ipx}, \, \, \la > 0, \ee
regarded as elements of the space of tempered distributions, form
an equicontinuous  family. Thus there exist subsequences
$\la_\iota \rightarrow 0$
such that the corresponding functions
$W_{\la_\iota}$
converge in the sense of distributions,
\be \lim_{\iota} W_{\la_\iota} ( x ) = W ( x ). \ee
(In order not to overburden the notation, we do not indicate the
possible dependence of the limit $W$ on the choice of the subsequence
or the number $N$ in the normalization constant.) Making use of the
KMS-condition, one finds that also the permuted functions
$W_\la^{\pi} ( -x ) = {Z_\la}^{-1} \omega ( \ax ( A_\la ) A_\la^* )$
converge for the respective subsequences $\la_\iota$,
\be \lim_{\iota} W_{\la_\iota}^{\pi} ( -x ) = W^{\pi} ( -x ). \ee
By performing this procedure for all possible sequences of operators
$ A_\la$ and numbers $N$ one arrives at a collection of tempered
distributions
$W, W^{\pi}$
which describe the correlations between arbitrary point-like
operations in the underlying KMS-state
$\omega$.
These distributions can be shown to have the following general
properties:

\nin i) {\em Locality:}
$W ( x ) = W^{\pi} ( -x )$
for spacelike
$x$.

\nin ii) {\em KMS-condition:} Denoting by $t$ the translations along the
time axis fixed by the rest system of the underlying heat bath at
temperature
${\beta}^{-1} > 0$,
the distribution-valued function
$t \rightarrow W ( x + t )$
has an analytic continuation into the strip
$ 0 < \mbox{Im} t < \beta$,
and its boundary value at the upper rim of this strip is given by
$\lim_{t \nearrow \beta} W ( x + i t ) = W^{\pi} ( -x ) $.

\nin iii) {\em Cluster-property:}
$ \lim_{t \rightarrow \pm \infty} W ( x + t ) = 0 $.

It turns out that there exists a universal integral representation
for these distributiones \cite{BB2}. We give here the result for the
case where the commutator function
$  C ( x ) = W ( x ) - W^{\pi} ( -x ) $
is antisymmetric in the time variable. (The general case can be reduced
to this one.)

\vspace{2mm}
\nin {\bf Proposition 5.1:} {\em Let $W$ be a tempered distribution
with
properties specified above. There holds
\be W ( x ) = \int_{0}^{\infty} dm \, D ( {\sx} ; m )
W^{(0)} ( x ; m ), \ee
where
$D ( {\sx} ; m ) $
is a distribution with respect to
${\sx}, m$
and
$W^{(0)} ( x ; m )$
is the correlation function of a non-interacting particle of mass $m$
in a thermal equilibrium state at temperature
${\beta}^{-1}$,
$$ W^{(0)} ( x ; m ) =
const \int dp \, \varepsilon ( p_0 ) \delta ( p^2 - m^2 ) \,
( 1 - e^{ - \beta p_0} )^{-1} {e^{ipx}}. $$}

This result provides us with the desired tool to uncover the
thermoparticle content of equilibrium states: the individual terms
$  D ( {\sx} ; m ) W^{(0)} ( x ; m ) $
in (25) describe the propagation of an excitation of mass $m$ which
has been created at the origin. Here the free correlation function
$W^{(0)} ( x ; m )$
corresponds to the unperturbed motion; it is in general modified by a
"damping factor"
$  D ( {\sx} ; m ) $
which subsumes the loss of probability of finding the excitation at
other space-time
points due to intermediate collisions. In the case of the stable
constituents of thermal equilibrium states, one would expect that
such collisions are the only mechanism by which these probabilities
are reduced, in contrast to excitations such as
metastable bound states, pairs and other composite systems, which
disintegrate also without external impact. We therefore conjecture
that thermoparticles contribute to the discrete (atomic) part of
the integral representations (25), whereas all other excitations
contribute to the continuous part. The contributions due to
thermoparticles can thus be
uncovered by inverting relation (25), cf. \cite{BB1}.

If the underlying thermoparticles are not "neutral", i.e. cannot be
created and annihilated by operators
$A \in \Al$,
one has to proceed from the algebra of observables to a corresponding
field algebra
$\cal F$
of charge-carrying fields. In the case of localizable charges, this
field-algebra is still generated by a net
$\Oo \rightarrow {\cal F} ( \Oo )$
of local field algebras \cite{DR}, and the thermal equilibrium states
can be extended to this net \cite{AHKT}. One can then proceed in
the same way as in the preceding discussion and construct a family
of correlation functions $W$, describing the effects of point-like
localized operations induced by fields in
$\cal F$.
Again there hold integral representations for the distributions $W$
of the kind given in relation (25), and the thermoparticles should
manifest themselves by discrete contributions in these
representations. This leads us to propose as a

\vspace{2mm}
\nin {\em Working hypothesis:} Thermoparticles, appearing as
constituents
of a thermal equilibrium state and carrying a localizable charge,
manifest themselves by discrete contributions in the integral
representations of the respective correlation functions of
point-like observables, respectively fields, cf. relations (23)
and (25).

\vspace{2mm}
In contrast to other proposals \cite{La}, our approach provides a
characterization of thermoparticles in unambiguous mathematical terms.
But there remain several open problems, and we conclude this section
by listing some of them.

i) What are the properties of thermoparticles? We have
identified
thermoparticles by the discrete contributions appearing in the
integral decomposition of the correlation functions $W$. In this way
we are able to specify their mass $m$ and the corresponding "damping
factors"
$  D ( {\sx} ; m ) $
which describe the dissipative effects of the thermal background.
Since these factors are responsible for the modification of the
dispersion law
of thermoparticles, it would be desirable to gather information about
their specific properties. For example, if the correlation functions
satisfy the relativistic KMS-condition proposed in \cite{BB3}, the
damping factors
$  D ( {\sx} ; m ) $
are locally analytic in {\sx}, which in turn implies that the
energy-momentum content of the corresponding thermoparticles is
concentrated about the mass shell
$p^2 = m^2$.
Another problem of interest is the determination of
the internal degrees of freedom of thermoparticles.

ii) How does one identify thermoparticles carrying
non-localizable charges? A solution of this problem does not seem
possible along the lines indicated in this section, since
such particles cannot be created by almost point-like operations in
view of the long-range fields which they carry along. Instead one
should try to localize the theromoparticles already present in the
state with the help of suitable localizing operators, in
analogy to the strategy outlined in Sect. 3.
By proceeding to point-like localizing operators, one
may then hope to extract
the thermoparticle content from the resulting correlation functions.

iii) How does one determine the thermoparticle content of spatially
non-homogenous states? The basic strategy should be the same as for
homogenous states, but there arise additional mathematical
complications in the construction and analysis of the correlation
functions $W$.

iv) When does a theory describe thermoparticles?
A general
analysis of this problem would require the following steps: first
one has to understand under which conditions the limits $W$ of the
correlation functions are non-trivial. This question is closely
related to the existence problem of point-like localized quantum fields.
In a second step one would then have to
clarify when there appear discrete contributions in the respective
integral representations of the distributions $W$. Here the conditions
of primitive causality and of causal independence, mentioned in
Sect. 3 in a similar context, should again be of relevance.

\section{Concluding ruminations}

The topic of the present article has been a conceptual problem in
quantum field theory: the notion of particle. We have seen that,
depending on the physical situation under consideration, the
description of particles leads to specific idealizations which call
for their own mathematical framework (particle weights, scaling
algebras, etc.).

The clarification of physical concepts requires a model independent
setting, and algebraic quantum field theory has proven to be a very
useful tool in this respect. As the motto of this workshop is an
invitation to contemplate about the achievements and perspectives of
our subjects, we conclude this article with some further remarks
and views about the value of model independent investigations for the
development and understanding of relativistic quantum field theory.

The idea that one may gain some insight into physics by studying a
mathematical setting (in contrast to a specific model) is sometimes
regarded with suspicion. It is argued that a general framework, such
as that of algebraic quantum field theory outlined in Sect. 2, contains
too little specific information and can therefore only yield "soft"
results which are applicable to all possible theories, no matter how
weird.

The point missed here is the fact that a general framework is the
appropriate arena for the {\em classification} of theories, and
detailed information can
frequently be obtained for the resulting
special cases. Let us mention as a prominent example the analysis of
Doplicher, Haag and Roberts, which led to a complete classification
of the possible statistics of states carrying a localizable charge
and the clarification of the relation between the type of statistics
and the structure of the underlying global gauge group.

The problem of classifying theories often leads also to new and
powerful {\em concepts}, an example being the "split property", first
conceived by Borchers, which proved to be useful not only in the
structural analysis but also in the investigation of concrete models.

A model independent analysis is furthermore the basis for the
development of general {\em algorithms}, well known examples being the
Haag-Ruelle collision theory and the Lehmann-Symanzik-Zimmermann
reduction formulas for the calculation of S-matrix elements. It also
leads to the formulation of {\em criteria} which
characterize physically significant
properties of a theory. For example, the analysis of the possible
range of the local degrees of freedom of quantum field theories,
initiated by Haag and Swieca, led to compactness and nuclearity
conditions which turned out to be a key e.g. to the understanding
of thermodynamical properties of quantum field theories.

Finally, the results of a model independent analysis may point
to the existence of interesting {\em new structures} which are
compatible with the general principles, but remained
undiscovered in the realm of models. An example of this kind are
theories of massive particles carrying a quantum topological charge.
Their specific properties were uncovered by Fredenhagen and the
present author in a general analysis, which subsequently
triggered a (successful) search for models.
(Cf. \cite{H} for more detailed information on the various results
mentioned.)

As may be seen from these remarks, it is one of the principle and
most interesting aims of algebraic quantum field theory to amend
the basic postulates by further conditions, resulting either from
a structural analysis or from physical considerations. The strategy
is thus to tighten the framework and thereby to obtain more detailed
information about the properties of physically significant theories.
One may even hope that this strategy leads to new methods for the
actual construction of models. Some interesting results in this
direction have recently been obtained by Wiesbrock \cite{Wi}, who
proved that conformal theories on the light ray can be constructed
from special inclusions of two von Neumann algebras.

Another objection raised against the general framework of (algebraic)
quantum field theory is the view that it may be too restrictive:
all attempts to construct a mathematically consistent and physically
significant quantum field theory fitting into this framework have
failed so far, and it may be necessary to include "gravitation" in
order to cure the agonizing ultraviolet problems. (The reader may
recognize this view in some of the contributions to these proceedings.)

It is difficult to argue against this kind of critique if one
cannot exhibit a counter example. But we dare to recall that the
gravitational constant is small and that the development of physics
has shown that "orders of magnitude" provide a natural grading of
phenomena. It would therefore be a surprise if the theoretical
understanding of present day high energy physics (and may be that
of the next century) could not be based in a consistent way on
the idealizations of special relativity. At least, there is
no indication that there exists some basic obstruction to
this effect. The present lack of relevant examples seems to have a
much simpler explanation: it is a difficult task to construct them.

In order to get an idea about the degree of difficulties it may be
instructive to compare the progress in quantum field theory with
the development of quantum mechanics. It took 30 years from Planck's
first observations till von Neumann's final rigorous formulation
of the theory. Vital intermediate steps were taken by Einstein,
Bohr, de Broglie and, finally, by Heisenberg, respectively
Schr\"odinger. So, roughly speaking, the time scale for progress in
quantum mechanics was 6 years.

The corresponding scale in quantum field theory seems to be about
20 years: quantum field theory was already conceived by the fathers
of quantum mechanics. The next vital step (covariant perturbation
theory and renormalization) was taken 20 years later by Tomonaga,
Schwinger, Feynman and Dyson and led to the consolidation of quantum
electrodynamics. It then took more than 20 years until the systematic
study of the ultraviolet properties of quantum field theories led
Gross and Wilczek, Politzer, Symanzik and others to the fundamental
concept of asymptotic freedom, which is now a basic ingredient in
the treatment of gauge theories, such as quantum chromodynamics.
According to this time scale some further important progress in
quantum field theory seems to be due, but one should not expect
that it will lead to its ultimate consolidation. It may take another
20 or 40 years until quantum field theory reaches a form
which also pleases mathematical physicists. So many of us may not
be able to appreciate it.

It is perhaps this unpleasant perspective which makes us susceptible
to all kinds of unconventional ideas which offer an apparently
quicker solution. But the solution of a difficult problem needs
time, sometimes much time, and continuous efforts (compare
mathematics). It would therefore be regrettable if
the beauties of physics in 2 and 10 dimensions let us forget the
inherent possibilities of the more conventional approach and
distract courageous people from pushing
this problem towards its solution.

But what if the lord is malicious and has enticed us to explore
for more than six decades some cul-de-sac? Probably the only way
to find him out would be a structural analysis which either reveals
the internal inconsistency of our present physical ideas or an
unavoidable clash between experimental facts and theoretical
predictions.
Thus the model independent analysis of the "basic principles" is a
vital part of the development of the theory and one may therefore
expect that algebraic quantum field theory will continue to play
its part in the consolidation of relativistic quantum
field theory also in the 21st century.
\newpage
\end{document}